\documentclass[sigplan,nonacm]{configs/acmart}
\makeatletter
\def\@ACM@checkaffil{
    \if@ACM@instpresent\else
    \ClassWarningNoLine{\@classname}{No institution present for an affiliation}%
    \fi
    \if@ACM@citypresent\else
    \ClassWarningNoLine{\@classname}{No city present for an affiliation}%
    \fi
    \if@ACM@countrypresent\else
        \ClassWarningNoLine{\@classname}{No country present for an affiliation}%
    \fi
}
\makeatother

\AtBeginDocument{%
  }

\setcopyright{none}
\copyrightyear{2025}
\acmYear{2025}
\acmDOI{XXXXXXX.XXXXXXX}

\usepackage{xcolor}
\usepackage{enumitem}
\usepackage{subfig}
\usepackage{adjustbox}
\usepackage{makecell}
\usepackage{tikz}
\usepackage{algorithm}
\usepackage{algpseudocode}
\usepackage{ulem}

\newcommand{\mao}[1]{\begingroup\color{cyan}#1\endgroup}
\newcommand{\name}{{\textsc{HybridGen}}}

\newcommand*\bcircled[1]{\tikz[baseline=(char.base)]{\node[circle, fill=black, inner sep=0.1ex, text=white] (char) {\scalebox{0.95}{#1}};}}

\begin{document}

\title{\name: Efficient LLM Generative Inference via CPU-GPU Hybrid Computing}

\author{Mao Lin}
\email{mlin59@ucmerced.edu}
\orcid{1234-5678-9012}
\affiliation{%
  \institution{University of California, Merced}
  \city{Merced}
  \state{California}
  \country{USA}
}

\author{Xi Wang}
\email{swang166@ucmerced.edu}
\affiliation{%
  \institution{University of California, Merced}
  \city{Merced}
  \state{California}
  \country{USA}
}

\author{Guilherme Cox}
\email{gcox@nvidia.com}
\affiliation{%
  \institution{NVIDIA}
  \city{Santa Clara}
  \state{California}
  \country{USA}
}

\author{Dong Li}
\email{dli35@ucmerced.edu}
\affiliation{%
  \institution{University of California, Merced}
  \city{Merced}
  \state{California}
  \country{USA}
}

\author{Hyeran Jeon}
\email{hjeon7@ucmerced.edu}
\affiliation{%
  \institution{University of California, Merced}
  \city{Merced}
  \state{California}
  \country{USA}
}


\begin{abstract}
As modern LLMs support thousands to millions of tokens, KV caches grow to hundreds of gigabytes, stressing memory capacity and bandwidth. Existing solutions, such as KV cache pruning and offloading, alleviate these but underutilize hardware by relying solely on either GPU or CPU for attention computing, and considering yet limited CPU local memory for KV cache storage. 
We propose \name, an efficient hybrid attention framework for long-context LLM inference.
\name\ enables CPU–GPU collaborative attention on systems with expanded tiered memory (e.g., CXL memory), addressing three key challenges: (1) multi-dimensional attention dependencies, (2) intensifying CPU-GPU load imbalance with longer sequences, and (3) NUMA penalty of tiered memories. \name\ tackles these by introducing attention logit parallelism, a feedback-driven scheduler, and semantic-aware KV cache mapping. Experiments with three LLM models with eleven different sizes on three GPU platforms with a CXL-expanded memory show that \name\ outperforms six state-of-the-art KV cache management methods by 1.41$\times$--3.2$\times$ on average while maintaining superior accuracy.


\end{abstract}


\received{20 February 2024}
\received[revised]{12 March 2024}
\received[accepted]{5 June 2024}

\maketitle

\section{Introduction}
\begin{figure}[t]
\centering
\includegraphics[width=0.9\linewidth]{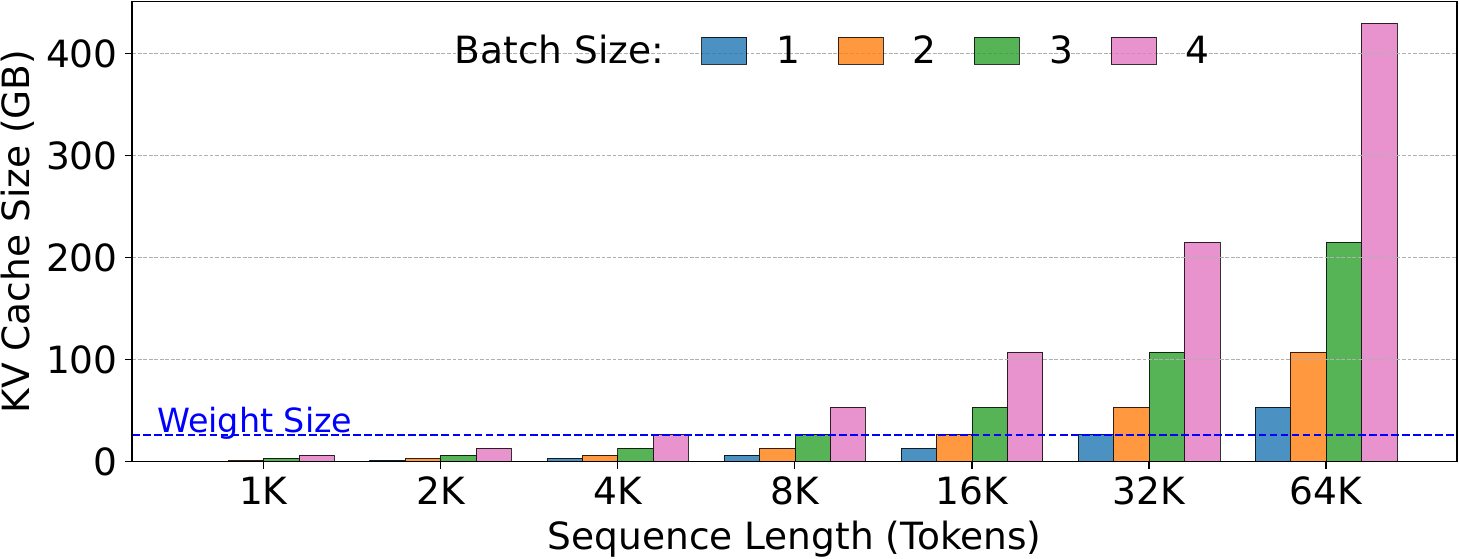}
\vspace{-5pt}
\caption{KV cache memory consumption of OPT-13B across varying sequence lengths and batch sizes. The dashed line indicates the model's weight size for comparison.}
\Description{}
\label{fig:kv_sizes_opt}\vspace{-5pt}
\end{figure}

Large language models (LLMs) have rapidly expanded across domains, with context lengths growing from GPT-1’s 512 tokens to millions in recent models like GPT-4.1 and Gemini 3 Pro.
As sequence length increases, key-value (KV) caches grow proportionally, significantly increasing memory usage and often exceeding GPU memory capacity.
For example, a 70B-parameter model with a 1M-token context can require hundreds of gigabytes of KV cache, 
making it a major bottleneck in LLM inference. Figure~\ref{fig:kv_sizes_opt} shows the escalating memory demand of the KV cache in the OPT-13B model~\cite{zhang2022opt}. 
Unlike fixed model weights (blue dotted line), the KV cache scales linearly with sequence length and batch size.

To address the KV cache size exceeding GPU memory, prior studies proposed KV cache pruning~\cite{kim2022learned, ge2023model,  liu2024minicache} and offloading~\cite{jiang2025kvpr, infinigen2024lee, sheng2023flexgen}. KV cache pruning retains only important tokens in GPU memory but risks accuracy loss from discarded context.  
Offloading preserves all KV tokens in the larger CPU memory to avoid accuracy degradation. However, GPU-based attention computing requires streaming KV entries back to the GPU for computation, incurring high data transfer overhead.
Recent studies mitigate this by selectively fetching important tokens~\cite{infinigen2024lee} or by offloading the entire attention computation to the CPU~\cite{moelightning,sheng2023flexgen}. Although these approaches reduce memory and bandwidth pressure on the GPU, they still fail to fully utilize available system resources.

Existing offloading approaches rely entirely on either GPU or CPU for attention computation, underutilizing system resources. Using only CPU memory wastes CPU compute power, especially as some modern systems adopt 1:1 or 1:2 CPU–GPU ratios (e.g., GH100 and GB200 NVL72~\cite{nvl72} platforms). On the other hand, fully offloading attention computation to the CPU unnecessarily wastes GPUs’ massive parallelism.
Furthermore, as KV cache sizes continue to grow, larger memory tiers such as CXL memory should be considered.
However, the performance implications of tiered memory for KV cache management have not been explored thoroughly~\cite{kim2025scalable, yoon2025tract,ipdps25:cxl,sc24_cxl}. A recent study~\cite{quinn2025longsight} examined KV cache management with CXL memory, but it targets attention on processing-in-memory (PIM) and is not applicable to conventional CPU–GPU systems.

\begin{figure}[t]
\centering
\includegraphics[width=1\linewidth]{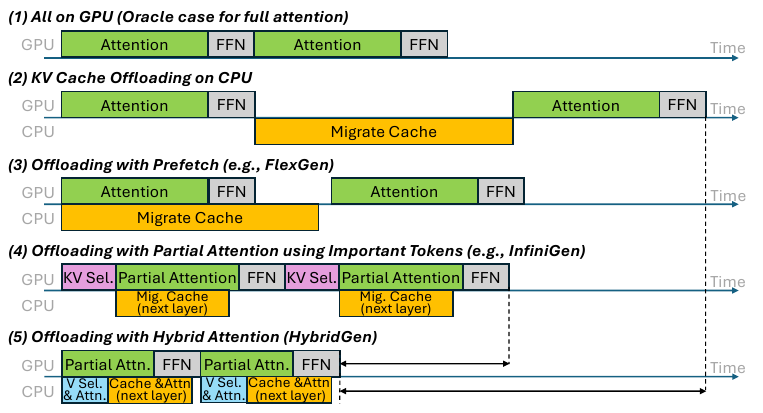}
\vspace{-5pt}
\caption{Estimated LLM inference time under various KV cache management: 
(1) Oracle case where GPU memory is large enough and attention can be done on GPU only, (2) Conventional KV cache offloading to CPU memory, where the KV caches are streamed to GPU from CPU memory everytime, (3) Advanced offloading where KV caches are prefetched to save migration overhead~\cite{sheng2023flexgen}, (4) State-of-the-art using selective attention where GPU runs an algorithm to identify important tokens and and fetch only those from CPU memory~\cite{infinigen2024lee}, and (5) \name, which parallelizes the attention computation on CPU and GPU, where both processors use tokens in their local memory, their workloads are balanced via a feedback scheduler, and KV storage is expanded to tiered memory via novel KV cache mapping.} 
\Description{}
\label{fig:timeline}\vspace{-5pt}
\end{figure}

In this paper, we introduce an efficient hybrid attention framework, \textit{\name}. \name\ parallelizes attention across CPU and GPU to fully utilize available compute power and exploit fast access to KV tokens in each processor’s local memory (e.g., GPU processes recent tokens, while CPU handles older offloaded tokens).
As shown in Figure~\ref{fig:timeline}, \name\ significantly reduces inference time compared to existing offloading and selective attention methods. 

To enable efficient hybrid attention computing, there are three challenges. \textbf{1) Multi-dimensional dependencies:} the attention pipeline has two strong dependencies, within and across layers.
Within a layer, softmax 
requires all $\mathbf{K}$ vectors, making it hard for CPU and GPU to concurrently process distinct sets of KV tokens. 
Across layers, since each layer’s output becomes the input of the next layer, the CPU must wait for the GPU to complete the remaining functions of the transformer block, limiting pipelined CPU-GPU execution.
To tackle this, \name\ decouples attention logits from the attention pipeline and enables the CPU to proactively compute the next layer’s attention using the current layer’s input, leveraging similarities between the inputs of consecutive transformer layers~\cite{moelightning, infinigen2024lee, liu2023deja, ying2021lazyformerselfattentionlazy}, thereby overlapping CPU–GPU computation.
\textbf{2) Inherent load imbalance between CPU and GPU:} 
As sequence length grows, due to limited GPU memory, more tokens will be offloaded to the CPU. Without a load-balancing mechanism, CPU will become the performance bottleneck. \name\ addresses this with a \textit{feedback scheduler} that dynamically adjusts CPU workload based on performance and accuracy.
\textbf{3) non-uniform-memory-access (NUMA) penalty of tiered memory:} To support large KV caches, we expand memory with tiered memory (e.g., via CXL).  
However, tiered memories are slower than local DRAMs~\cite{weisgut2025vldb, eejournal2025}. To mitigate this NUMA impact, \name\ proposes a \textit{semantic-aware KV cache mapping} that removes CXL access latency from the critical path.

Our contributions are as follows: 
\begin{enumerate}[label=\textbullet, leftmargin=*, labelindent=2pt]
\item We provide an in-depth analysis of existing KV management methods and show potential benefits and challenges of hybrid attention computing for 
LLM inference. \name\ 
realizes well-balanced pipelined hybrid attention computing with lightweight software modules on off-the-shelf CPU-GPU platforms.

\item To efficiently orchestrate hybrid attention computing, we propose a feedback scheduler.
The feedback scheduler uses both 
performance and accuracy statistics for balancing the loads between CPU and GPU, unlike existing solutions 
\textcolor{black}{that focus on either accuracy or performance, not both}. 

\item To accommodate increasing KV cache size, \name\ supports emerging tiered memory architecture. 
To alleviate the NUMA impact of 
tiered memory, we propose a semantic-aware KV cache mapping, which effectively removes long memory access time from the critical path.


\item We evaluate \name\ 
and six state-of-the-art KV cache management baselines with diverse LLMs in various sizes on three GPU platforms, with and without a CXL shared memory pool. 
\name\ outperforms the compared solutions by 1.41$\times$--3.2$\times$ on average with superior accuracy.

\end{enumerate}

\section{Background}\label{sec:background}

\subsection{LLM Generative Inference}

\begin{figure}[t]
\centering
\includegraphics[width=0.95\linewidth]{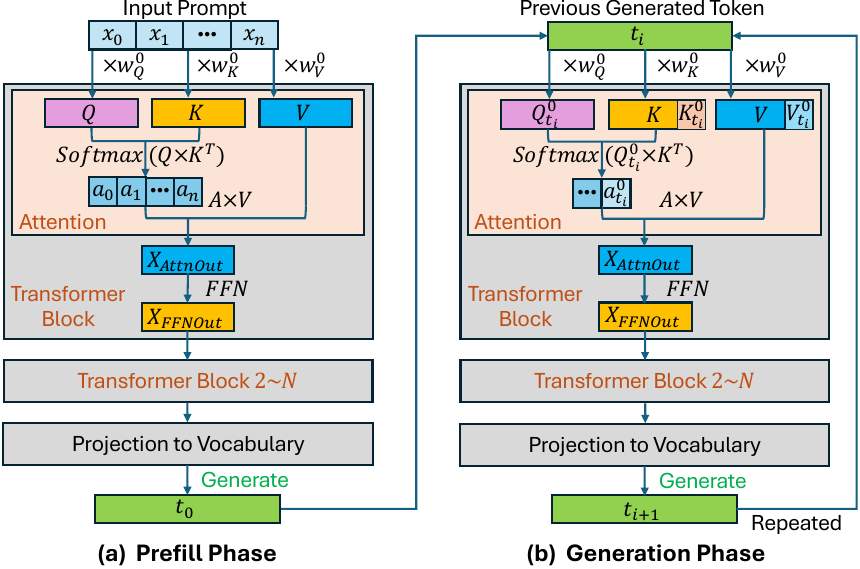}
\vspace{-5pt}
\caption{LLM generative inference.}
\Description{}
\label{fig:llm_inference}\vspace{-5pt}
\end{figure}

LLM inference consists of two stages: \textit{prefill} and \textit{generation} (or \textit{decode}). Each token attends to all prior tokens in LLM inference, making recomputation of past key ($\mathbf{K}$) and value ($\mathbf{V}$) vectors prohibitively expensive.
Modern LLMs avoid this by caching $\mathbf{K}$ and $\mathbf{V}$ tensors from prefill and updating them incrementally during generation, known as the \textit{KV cache}.

During prefill (Figure~\ref{fig:llm_inference}(a)), 
the model processes the entire input prompt (e.g., $x_0, \ldots, x_n$) 
and produces the first output token ($t_0$), serving as the initial input to the subsequent generation iterations. 
For each transformer layer, key and value vectors of all tokens are computed and stored in the KV cache for reuse.
During generation, the model autoregressively produces tokens. As shown in Figure~\ref{fig:llm_inference}(b), for a new token $t_i$, the query, key, and value vectors ($\mathbf{Q}^0_{t_i}$, $\mathbf{K}^0_{t_i}$, $\mathbf{V}^0_{t_i}$) are computed at the first layer, and the new $\mathbf{K}^0_{t_i}$ and $\mathbf{V}^0_{t_i}$ are appended to the KV cache.

Attention logits are computed as $\mathbf{Q}^0_{t_i} \times \mathbf{K}^T / \sqrt{d}$, where $d$ is the per-head dimension. These logits capture raw relevance between the current and past tokens, where larger logits indicate stronger relevance. Applying \textit{softmax} produces attention scores that weight value vectors to generate the attention output, where higher scores indicate greater influence. This process continues until an end-of-sequence (EoS) token is produced or a maximum length is reached.

relevance between the current token $t_i$ and previous tokens, where larger logits indicate stronger relevance. 

\subsection{Long Context LLM Inference}
Long contexts, spanning hundreds of thousands to millions of tokens, 
enable LLMs to capture long-range dependencies and improve output quality. 
However, as sequence length (context) increases, the KV cache 
size also scales linearly, 
creating significant memory pressure~\cite{zhang2023h2o}.
Several studies tackled the exploding KV cache size through KV cache offloading and pruning. 
KV cache pruning retains only a subset of important tokens in GPU memory using strategies such as recent-window attention~\cite{xiao2024streaming, liu2023scissorhands}, attention sinks~\cite{xiao2024streaming, zhang2023h2o, li2024snapkv}, or fixed-size top-$K$ selection~\cite{zhang2023h2o, li2024snapkv, ghadia2025dialogue}. While these reduce memory usage and enable constant-time attention, they may degrade accuracy due to discarded context.
Offloading preserves all KV tokens by storing them in larger CPU memory or storage~\cite{infinigen2024lee, yu2025stateful, sheng2023flexgen} to avoid accuracy loss. However, it requires expensive KV token streaming back to the GPU at each generation step. 
To mitigate this, recent work proposes speculative prefetching of important tokens~\cite{infinigen2024lee}. It predicts important tokens by approximating attention using the prior layer’s input, exploiting the inherent similarity between inputs of consecutive layers.

\subsection{Memory Expansion via CXL}
As application size becomes bigger, memory subsystems have also evolved quickly.
One promising solution is memory expansion via Compute Express Link (CXL), an industry standard that enables connecting memory and compute devices via PCIe.
Through CXL protocol, a set of DRAMs or SSDs can be connected to PCIe switch, mapped to the host CPU address space, and accessed by the host CPU or other accelerators (e.g., GPUs) via memory load/store instructions. Due to the extra PCIe and protocol latencies, CXL memories are known to be a few times slower than local memories. 
Thus, several studies introduced data mapping and prefetching algorithms between CXL and local memories, considering data access frequency (hotness) and locality~\cite{kim2025lia, ipdps25:cxl, sun2023cxl, zhao2025optimizing}. 
\section{Motivation}
This section analyzes the benefits and limitations of existing KV cache offloading methods when attention computation is done either on CPU or GPU.

\subsection{Data Traffic Depending on Attention Affinity}
\label{sec:mov_data_traffic}
\begin{figure}[t]
\centering
\includegraphics[width=0.85\linewidth]{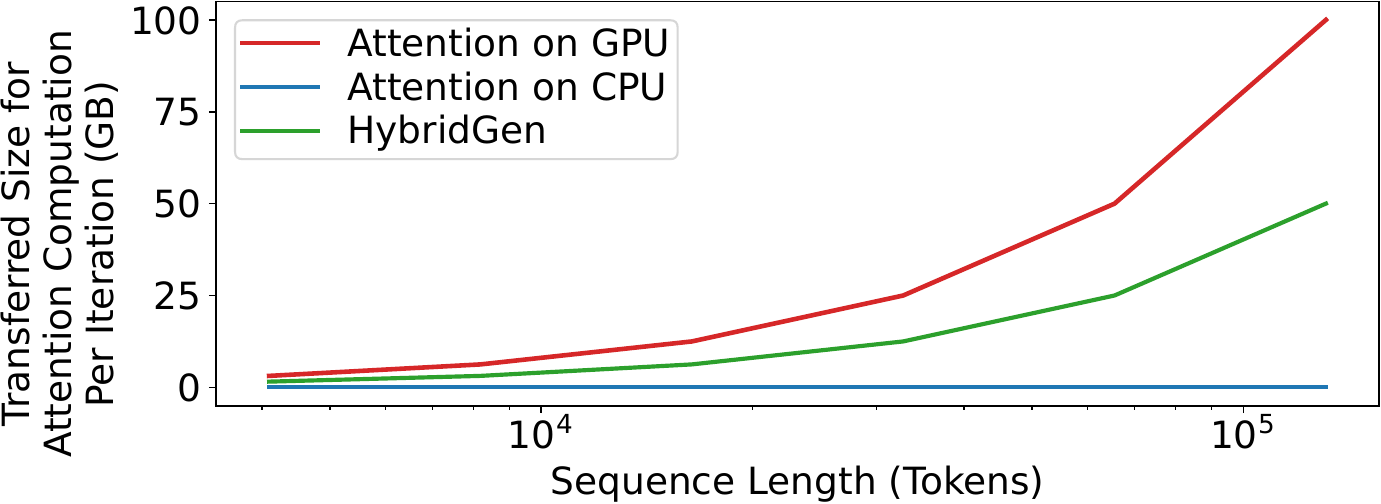}
\vspace{-5pt}
\caption{Estimated data traffic during attention layer computation per iteration under different strategies. }
\Description{}
\label{fig:weight_offloading}\vspace{-5pt}
\end{figure}

\begin{figure}[t]
\centering
\includegraphics[width=0.85\linewidth]{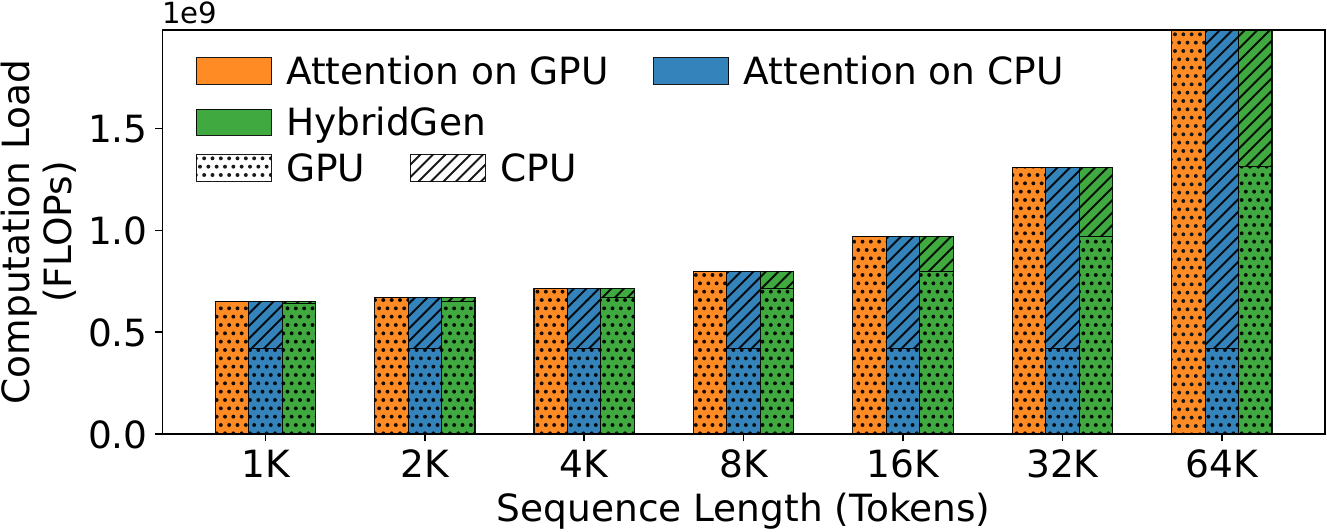}
\vspace{-5pt}
\caption{Computation distribution on transformer block computation per iteration under different strategies.}
\Description{}
\label{fig:comp_dist}\vspace{-5pt}
\end{figure}

To analyze how processor affinity affects data transfer overhead, we measure data traffic under two settings: (1) KV cache offloaded to CPU with attention on GPU (\textit{AoG}), and (2) both KV cache and attention on CPU (\textit{AoC}).
In AoG, both key-value tensors must be transferred from CPU to GPU, resulting in $2 \times N \times d$ data transfer volume per layer, where $N$ is sequence length and $d$ is attention dimension. 
In contrast, AoC transfers only the fixed-size output $d$, incurring constant data traffic. 
Thus, AoG scales linearly with sequence length, while AoC remains constant, as shown in Figure~\ref{fig:weight_offloading}. 
As sequence length increases, AoC becomes more advantageous in reducing data transfer.

\subsection{Load Imbalance Depending on Attention Affinity}
Data traffic alone does not determine which approach is better, so we break down the computation loads of CPU and GPU when different sequence lengths are used. 
As shown in Figure~\ref{fig:comp_dist}, AoG exhibits the most imbalanced computation load across CPU and GPU, because GPU handles all computations, including attention, FFN, and layernorm, etc. Though GPU has a higher throughput than CPU, given the increasing amount of data when the sequence becomes longer, GPU can be easily overloaded, while leaving CPU idle. 
In contrast, AoC maintains nearly constant GPU computation load per decoding iteration, as non-attention operations are independent of sequence length.
However, CPU workload increases with longer sequences due to attention over larger KV caches. 
At 64K sequence length, CPU computes almost 5$\times$ more operations than GPU.
Since CPUs have lower parallelism, AoC is not scalable despite reducing data traffic compared to AoG.

\subsection{Ball-park Latency Estimation}\label{sec:motiv-challenge}

We also did a ballpark latency estimation of the two approaches by using an estimated theoretical FLOPS of CPU and GPU as 46 GFLOPS and 1.3 TFLOPS based on the machine A in Table~\ref{tab:platform}~\footnote{A fused multiply–add (FMA) operation is assumed to take 4 cycles on the NVIDIA A100 GPU \cite{abdelkhalik2022demystifying}, while a floating-point operation typically takes 5 cycles on the Intel CPU \cite{intel-latency}.
The NVIDIA A100 consists of 6912 CUDA cores running at 765 MHz, providing an approximate throughput of 1.3 × 10\textsuperscript{12} operations per second, whereas the Intel Xeon CPU has 104 cores at 2.2 GHz, yielding roughly 4.6 × 10\textsuperscript{10} operations per second.}.
The transfer time is measured by migrating a given amount of data using the memory-copy API on the machine.
We also assumed that all computations can find data from the local memory through perfect prefetching. As we do not assume any computation and data traffic overlapping scheme, we show the data transfer latency and computation latency in separate bar charts. As shown in Figure~\ref{fig:comp_balance}, AoG encounters higher data transfer latency while AoC suffers from at least 4$\times$ longer execution time than AoG due to CPU’s lower computational throughput.

\subsection{CPU-GPU Hybrid Attention Computing}
\begin{figure}[t]
\centering
\includegraphics[width=0.85\linewidth]{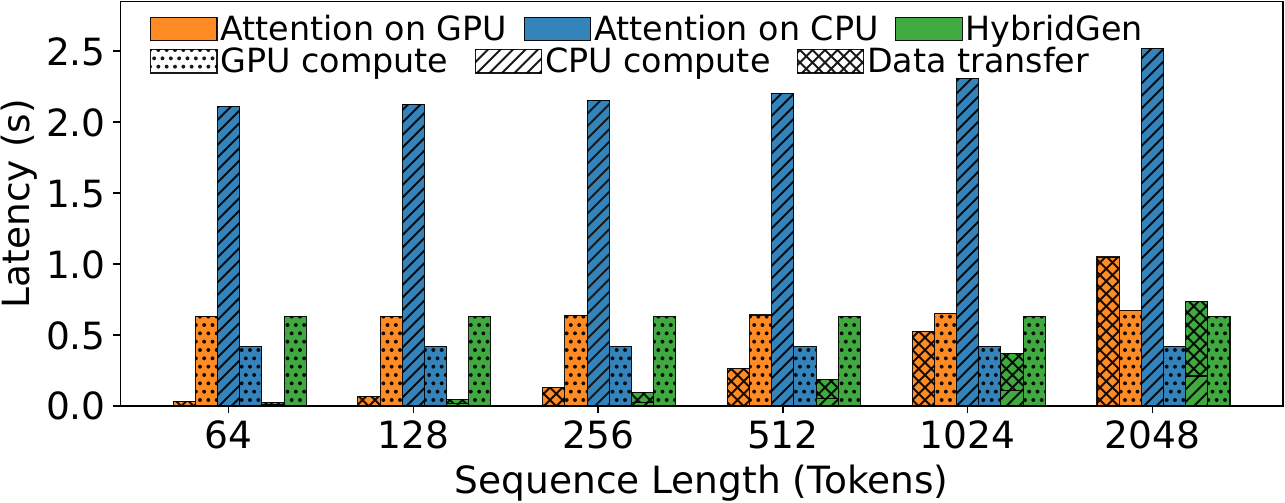}
\vspace{-5pt}
\caption{Estimated latency under different strategies. 
}
\Description{}
\label{fig:comp_balance}\vspace{-5pt}
\end{figure}

The previous results show obvious limitations of both solutions: AoG is constrained by memory capacity, while AoC is limited by compute capacity. This motivates CPU–GPU hybrid attention computing.
In this design, the CPU computes attention for tokens offloaded to CPU memory, while the GPU handles tokens residing in GPU memory (e.g., recently generated ones) and completes the remaining attention pipeline.
As shown in Figures~\ref{fig:weight_offloading} - 
\ref{fig:comp_balance}, hybrid attention computing (denoted as \name) reduces data transfer by avoiding movement of raw KV vectors, while leveraging GPU parallelism.
At the same time, as CPU only calculates attentions of a subset of tokens, we do not suffer from CPU's limited compute capacity. With these benefits, our estimated performance shows that hybrid computing is the best among the three approaches.



\begin{figure}[t]
\centering
\includegraphics[width=0.85\linewidth]{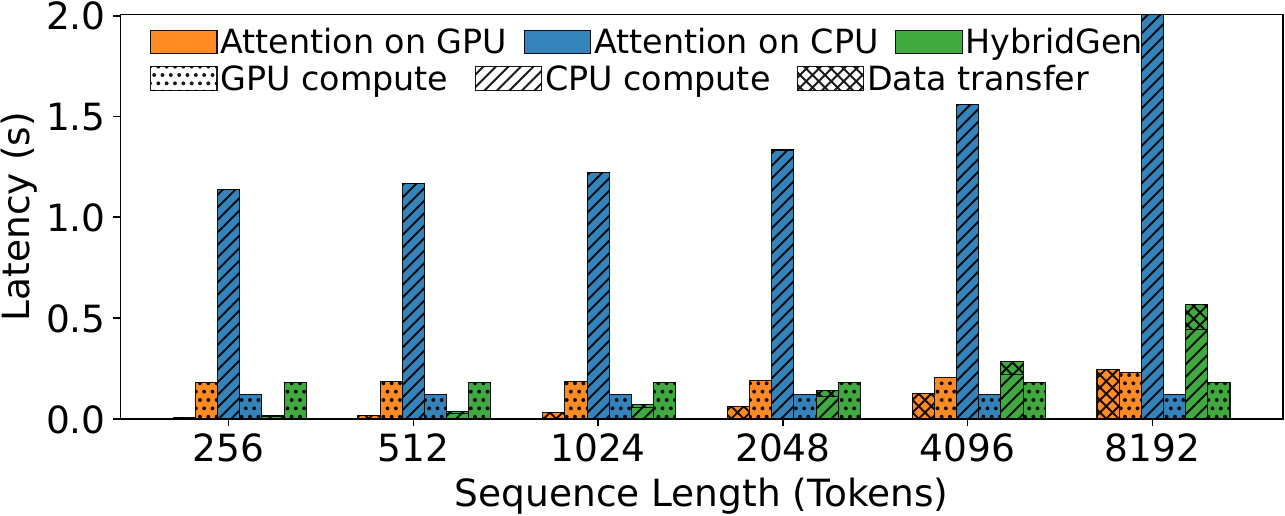}
\vspace{-5pt}
\caption{Estimated latency of Grace-Hopper-like architecture: \textit{Note that due to the limited GPU resource, AoG can not process longer sequences or larger batches (e.g., 3490 tokens with a batch size of 4), while the other two approaches can.}}
\Description{}
\label{fig:comp_balance_gh_full}\vspace{-5pt}
\end{figure}

\begin{figure}[t]
\centering
\subfloat[Layer 13.\label{fig:layer_36}]{
    \centering\includegraphics[width=0.42\linewidth]{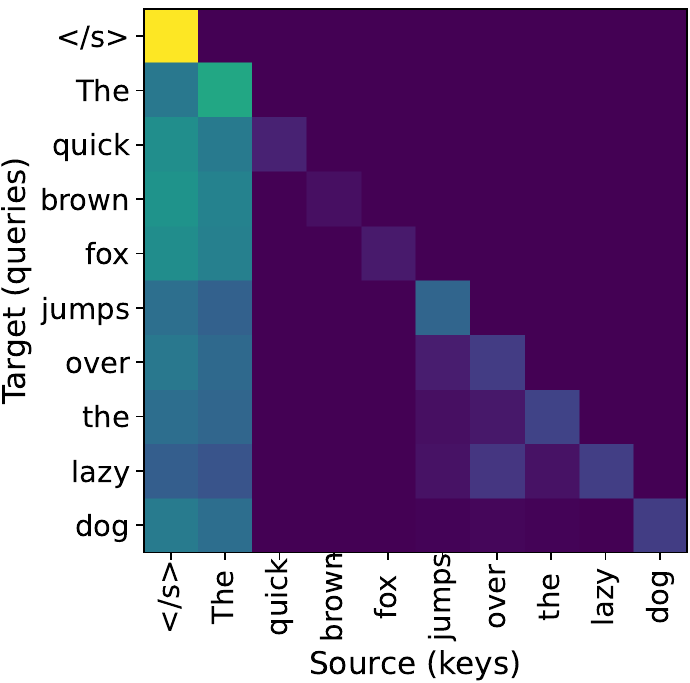}
}
\subfloat[Layer 36.\label{fig:layer_10}]{
    \centering\includegraphics[width=0.5\linewidth]{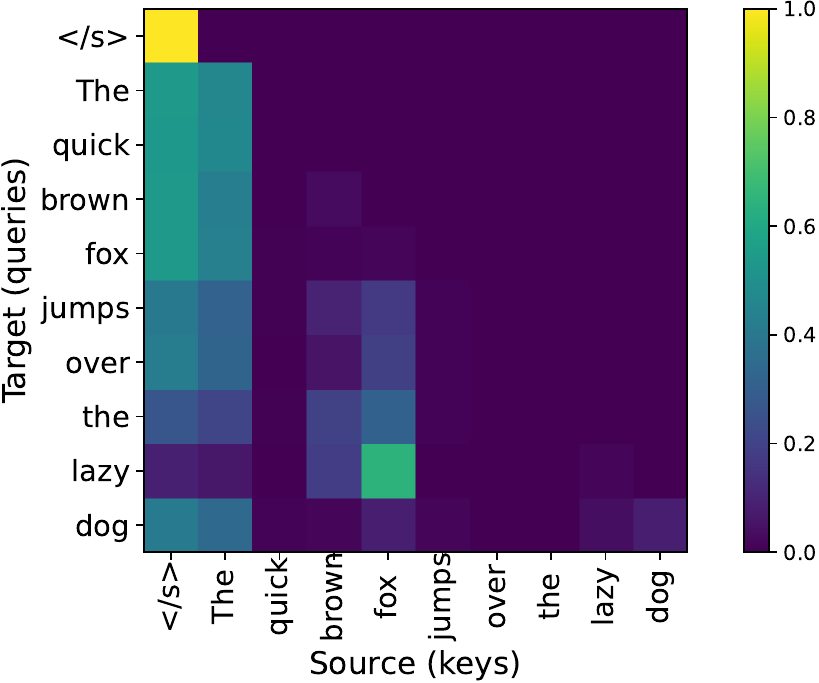}
}
\\\vspace{-5pt}
\caption{Heatmap of the attention scores from different layers in OPT-13B. }
\label{fig:attention_score}\vspace{-5pt}
\end{figure}

\begin{figure}[t]
\centering
\includegraphics[width=0.85\linewidth]{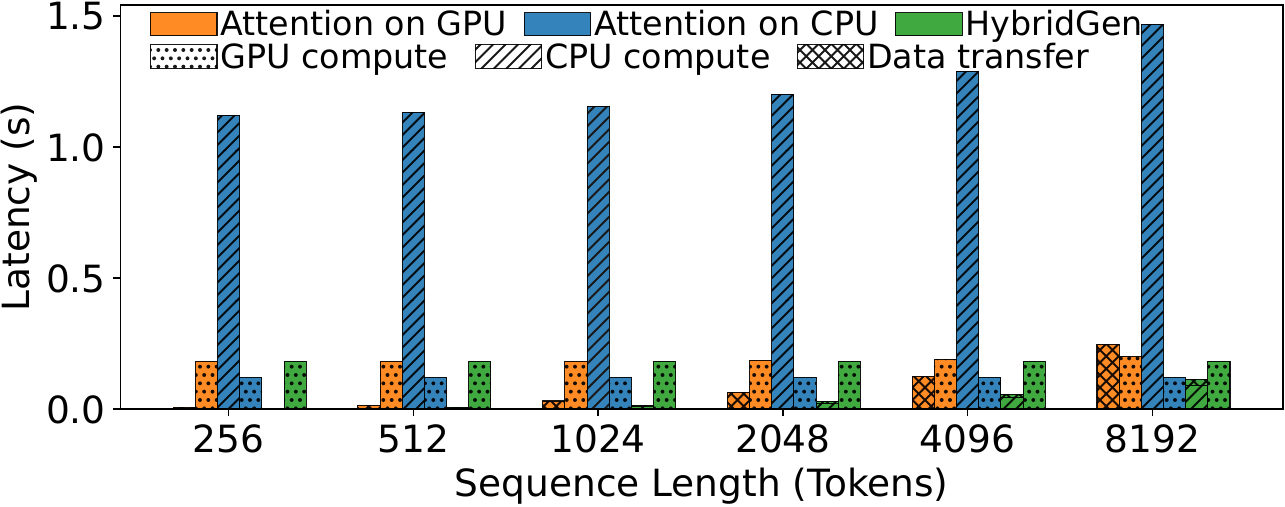}\vspace{-5pt}
\caption{Estimated latency of Grace-Hopper-like architecture with KV cache pruning: 40\% tokens used for attention.}
\Description{}
\label{fig:comp_balance_gh_selective}\vspace{-5pt}
\end{figure}

\subsubsection{Can Faster Interconnect Help?}
Recent superchip architectures (e.g., GH100) integrate CPU and GPU within a single package using high-bandwidth interconnects (NVLink C2C), significantly improving data transfer speed. 
To evaluate whether this reduces AoG’s data transfer overhead, we estimated latencies under the GH100 configuration~\cite{gh100}, as shown in Figure~\ref{fig:comp_balance_gh_full}.
Note that results are shown only when all three approaches can process the given sequence. AoG 
failed for longer sequences or larger batches 
due to limited GPU resources, while AoC and \name\ remain feasible.
Compared to Figure~\ref{fig:comp_balance}, the faster interconnect effectively reduces memory copy overhead. 
However, due to the limited parallel processing power and local DRAM bandwidth, AoC is the least desirable solution. \name\ achieves better performance than AoG, but AoG can surpass \name\ at longer sequence lengths. However, AoG is not scalable due to stricter limits on input size and batch size. 
Thus, faster interconnects mitigate data transfer overhead but cannot be an ultimate solution for long-context LLM inference. 

\subsubsection{Can KV Cache Pruning Help?}

As discussed in Section~\ref{sec:background}, KV cache pruning maintains only selected tokens, so it could solve the KV cache memory pressure problem. 
However, such methods risk accuracy loss by discarding \textit{seemingly} unimportant tokens~\cite{zhang2023h2o, xiao2024streaming, ghadia2025dialogue}.
This is because important token positions vary across iterations, layers, and heads, despite some common patterns.
To examine the consistency of important token positions, we measured attention scores across iterations for individual 
layers. 
Figure~\ref{fig:attention_score} shows a heatmap of the attention scores of two sample layers. 
Findings are: 
1) tokens important in one iteration (e.g., ``over'', ``lazy'' in iteration 9) may become unimportant in the next iteration, and 2) while early and recent tokens are often important, middle tokens can also be critical without a consistent pattern. 
Thus, for accuracy, 
KV cache offloading (i.e., preserving 
context) and dynamic token pruning 
are needed.

To analyze AoG, AoC, and \name\ under dynamic token selection, we re-estimated latency on GH100 assuming 40\% of KV tokens are important. For \name, we assign 20\% tokens to CPU and 20\% to GPU, excluding token prediction latency due to algorithm variability. Figure~\ref{fig:comp_balance_gh_selective} shows the results.
Dynamic token selection 
significantly improves performance by reducing token counts. However, AoC still scales poorly due to limited compute power. 
\name\ outperforms both approaches by balancing workloads between CPU and GPU and leveraging GPU parallelism for remaining computation. This demonstrates that, even with identical pruning, performance strongly depends on workload mapping strategy.




\section{\name{}}
\label{sec:hybridgen}

\subsection{Overview \& Challenges}

We propose \name, which tackles the computation and data traffic imbalances of AoC and AoG by leveraging both CPU and GPU resources. 
Unlike prior approaches, \name\ offloads a large portion of KV cache and attention computation to the CPU, enabling both CPU and GPU to operate on tokens in their local memory, thereby reducing data transfer and improving utilization.

However, there are several challenges for hybrid attention computing; 1) the attention pipeline has intra-layer and inter-layer dependencies, 
2) due to limited GPU memory capacity, it is hard to balance the loads; CPU 
needs to process more tokens for longer sequences, 
and 3) CPU memory capacity could also reach the limit for a significantly longer context. 

\begin{figure}[t]
\centering
\includegraphics[width=0.9\linewidth]{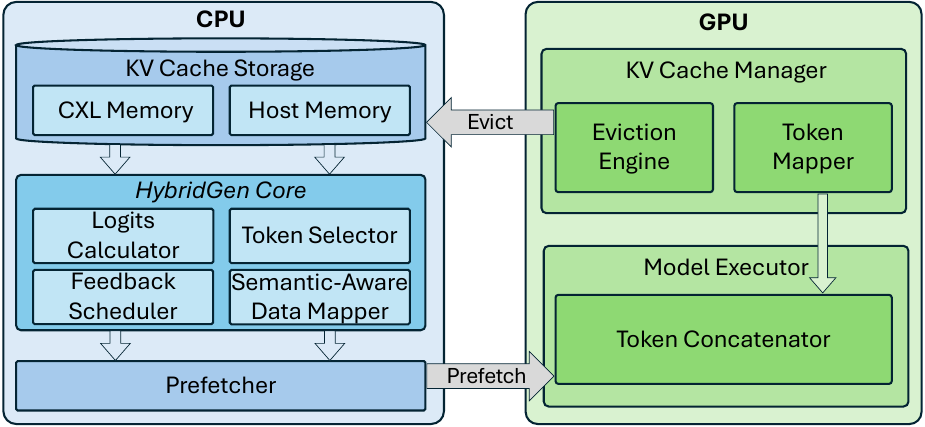}
\caption{Architecture of \name{}. 
}
\Description{}
\label{fig:design}\vspace{-5pt}
\end{figure}

\name\ addresses these challenges with a lightweight software framework, 
as shown in Figure~\ref{fig:design}. 
Section~\ref{sec:attn_parallelization} describes how \name\ resolves 
attention dependencies (challenge 1) with  \textit{Logits Calculator} on CPU and \textit{Token Concatenator} on GPU.
Section~\ref{sec:feedbacksched} presents a novel \textit{Feedback Scheduler} that dynamically balances CPU–GPU workloads as sequence length increases (challenge 2) to prevent CPU bottlenecks while preserving accuracy. 
Section~\ref{sec:semantic} describes how \name\ extends to shared memory systems (e.g., CXL) for super long sequences (challenge 3) via a \textit{Semantic-aware Data Mapper} that mitigates NUMA overhead.

\subsection{Attention Process Parallelization}
\label{sec:attn_parallelization}

\name\ aims to enable 
\textit{parallel attention computation across CPU and GPU}. 
However, 
partitioning attention computation across heterogeneous processors is 
challenging due to strong data dependencies in the attention pipeline. 
As illustrated in Figure~\ref{fig:llm_inference}, 
conventional attention first computes logits $\mathbf{Q}\mathbf{K}^T$ (scaled by $1/\sqrt{d}$), followed by softmax to obtain attention scores over \textit{all} tokens.
These attention scores capture the contextual relevance between the current token and historical tokens, and are subsequently used for value $\mathbf{V}$ aggregation.
This softmax operation introduces a global dependency across tokens, making partitioning difficult.
To address this, \name{} 
decouples attention computation into two stages and parallelizes the first stage.
Specifically, while the softmax and value aggregation steps require global logits, the \textit{attention logit computation} itself, $\mathbf{Q} \cdot \mathbf{K}^{T}$, is fully independent across tokens.
Exploiting this, \name\ allows the CPU and GPU to compute logits for keys in their local memory in parallel. 
The GPU then concatenates all logits, applies softmax, and completes value aggregation and the remaining transformer operations.


\begin{figure}[t]
\centering
\includegraphics[width=0.95\linewidth]{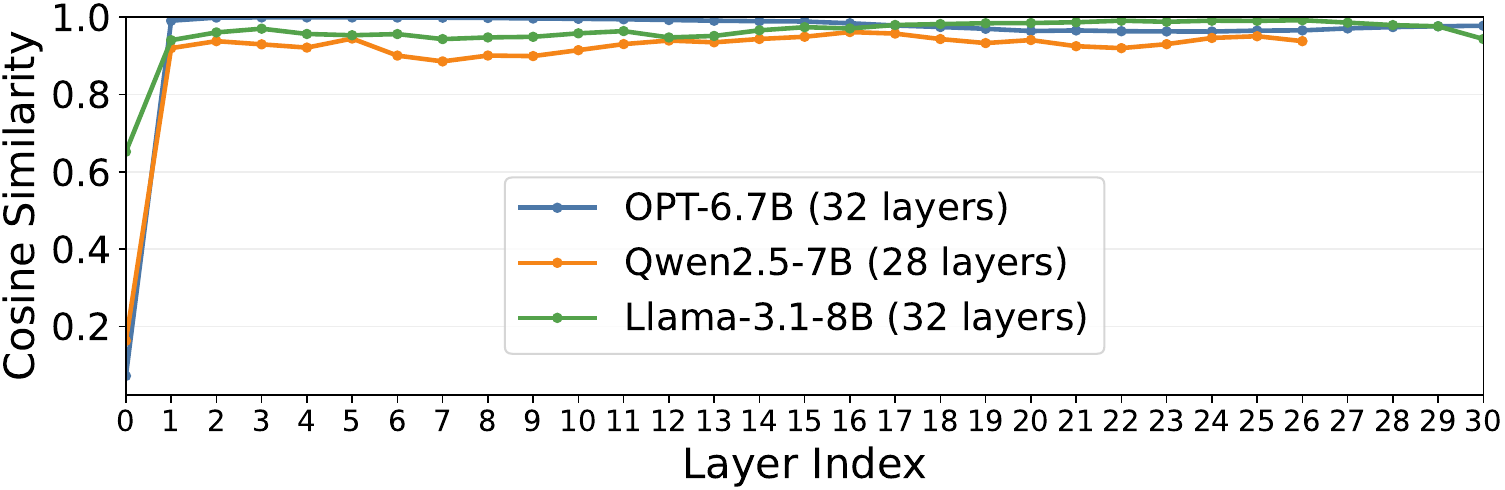}
\caption{Cosine similarity between inputs of consecutive transformer layers (layer i vs. layer i+1) during decoding for OPT‑6.7B, Qwen2.5‑7B, and Llama‑3.1‑8B.}
\Description{}
\label{fig:layer_cosine_similarity}\vspace{-5pt}
\end{figure}

\begin{figure*}[t]
\centering
\includegraphics[width=0.95\linewidth]{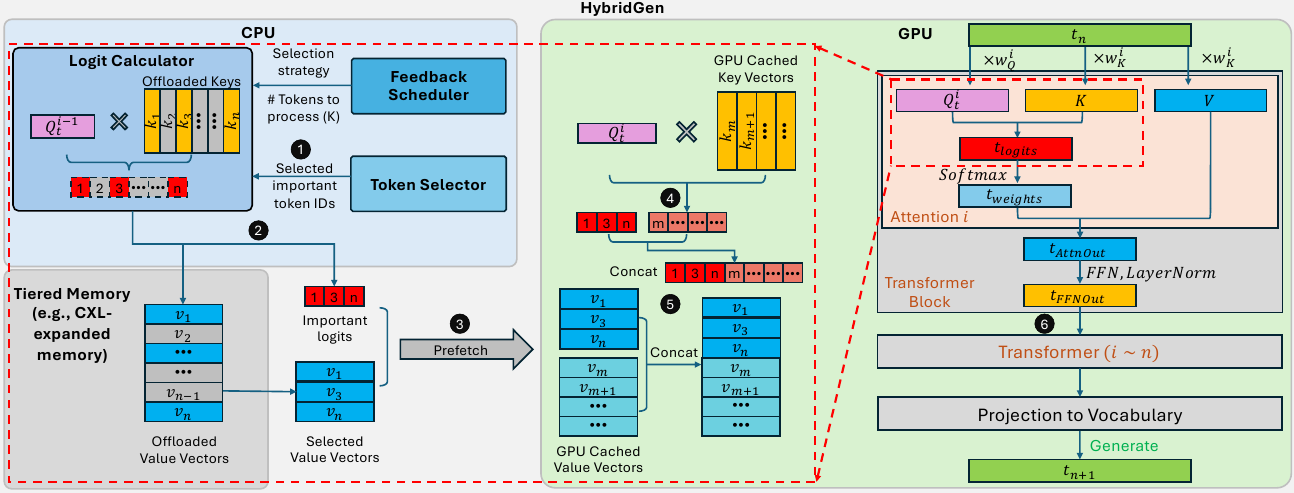}
\vspace{-5pt}
\caption{Workflow of \name{}: 
\textcircled{1}~While the GPU executes attention and the FFN for layer $i$, the CPU selects important tokens and computes attention logits for layer $i+1$ using the input of layer $i$, leveraging similarity across consecutive layers.
\textcircled{2}-\textcircled{3}~The CPU transfers the computed logits and corresponding value vectors from tiered memory to the GPU.
\textcircled{4} In parallel, the GPU computes logits for tokens cached in GPU memory.
\textcircled{5}~The GPU concatenates logits and value vectors, restores the original token order, applies softmax, performs value aggregation, and completes the remaining transformer computations.
\textcircled{6}~The GPU transfers the transformer block output to the CPU for the next layer.}
\Description{}
\label{fig:workflow}\vspace{-5pt}
\end{figure*}

Another dependency exists across layers: since the next layer’s input is produced by the GPU, the CPU must wait until the GPU completes the attention pipeline for each layer. 
To tackle this, we leverage similarity between the inputs of consecutive transformer layers.
As discussed in Section~\ref{sec:similarity} and observed in prior works~\cite{infinigen2024lee,ying2021lazyformerselfattentionlazy, moelightning}, 
layer inputs change gradually and are highly similar. Prior studies show that reusing inputs across consecutive layers (with layer-specific weights) yields comparable accuracy~\cite{ying2021lazyformerselfattentionlazy}. 
Exploiting this, the CPU proactively computes attention logits for the next layer using the current layer’s input and next-layer weights, while the GPU completes the current layer. The CPU results are then used by the GPU to finish attention in a pipelined manner, after which the output is shared back with the CPU.

\subsubsection{Consecutive‑Layer Input Similarity}
\label{sec:similarity}
To validate our key assumption for mitigating inter-layer dependencies, we measured the cosine similarity between inputs of consecutive transformer layers (the residual stream before each layer) on three LLMs, as shown in Figure~\ref{fig:layer_cosine_similarity}. 
We ran decoding on a common prompt set and calculated similarity across all consecutive layer pairs 
across the model depth. 
Results show that 
similarity increases sharply after the first layer and remains high throughout the network for all models. This indicates that neighboring layers operate on highly correlated representations, supporting the use of layer $i$'s input as a proxy for layer $i{+}1$ in our design.

\subsubsection{CPU-Side Support for Hybrid Attention}
To enable CPU-side attention logit computation, \name\ incorporates a \textit{Logits Calculator}, as shown in Figure~\ref{fig:design} and \ref{fig:workflow}. 
It computes attention logits using $\mathbf{K}$ vectors in CPU memory, after which the CPU transfers the resulting logits (a compact vector) and corresponding $\mathbf{V}$ vectors to the GPU.
By transferring logits and selected $\mathbf{V}$ vectors instead of raw $\mathbf{K}$ and $\mathbf{V}$ vectors, \name{} significantly reduces CPU--GPU data movement, as also discussed in Section~\ref{sec:mov_data_traffic}.

For a pipelined CPU--GPU computing, the logits calculator 
enables the CPU to compute attention logits for layer~$i\!+\!1$ using the input of layer~$i$ and the weights of layer~$i\!+\!1$, while the GPU executes attention and FFN computation for layer~$i$. 
This pipelined execution model effectively overlaps CPU and GPU computations, thereby improving overall utilization and reducing end-to-end inference latency.

\subsubsection{Device-Side Support for Hybrid Attention}
\label{sec:gpucomp}

\name\ employs a set of device-side components to preserve correct attention semantics under hybrid CPU--GPU execution.
These components ensure functional equivalence to conventional attention despite KV cache partitioning, without introducing any new algorithmic behavior or requiring hardware modifications.  
\textit{Eviction Engine} manages KV cache residency by offloading older tokens from GPU to CPU under memory pressure, using a least-recently-generated (LRG) replacement policy~\cite{infinigen2024lee, sheng2023flexgen}. Evicted tokens remain available for CPU-side attention computation. 
The lightweight \textit{Token Mapper} tracks positional ranges of GPU- and CPU-resident KV segments, allowing the GPU to reassemble attention logits without runtime sorting or synchronization.
\textit{Token Concatenator} merges CPU- and GPU-computed logits and value vectors in the original token order tracked by the token mapper. The GPU then applies softmax and completes value aggregation, producing outputs identical to monolithic execution while enabling parallel CPU--GPU computation.

\begin{figure}[t]
\centering
\includegraphics[width=0.85\linewidth]{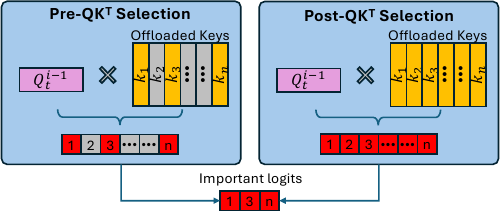}
\caption{Attention logits computation under two token selection mechanism: Feedback scheduler chooses one of these at runtime to balance the performance.
}
\Description{}
\label{fig:token_selection}\vspace{-5pt}
\end{figure}

\begin{algorithm}[t]
\small
\caption{Feedback Scheduler Policy}
\label{alg:scheduler}
\begin{algorithmic}[1]
\Require GPU stage latency $T_{\mathrm{gpu}}$
\Require CPU stage latency $T_{\mathrm{cpu}}$
\Require Transfer latency $T_{\mathrm{tx}}$
\Require Offline accuracy threshold $K_{\min}$
\Ensure Token count $K$, selection strategy

\State Initialize $K \gets K_{\min}$
\State Initialize strategy $\gets$ Post-QK$^{\mathrm{T}}$

\If{$T_{\mathrm{cpu}} + T_{\mathrm{tx}} \le T_{\mathrm{gpu}}$}
    \State Increase $K$ to improve accuracy
\Else
    \If{$K$ > $K_{\min}$}
        \State Decrease $K$ while ensuring $K \ge K_{\min}$
    \Else
        \State strategy $\gets$ Pre-QK$^{\mathrm{T}}$
    \EndIf
\EndIf


\State \Return $(K,\ \text{strategy})$
\end{algorithmic}
\end{algorithm}

\subsection{Load Balancing with Feedback Scheduler}
\label{sec:feedbacksched}

In hybrid attention, effectively balancing CPU and GPU work is critical for high utilization and low latency. As sequence length grows, more KV tokens are offloaded to the CPU, increasing CPU-side computation and CPU--GPU data transfer costs and potentially making the CPU the bottleneck. To address this challenge, \name{} introduces a \textit{Feedback Scheduler} that dynamically adjusts the CPU-side workload based on runtime latency feedback while enforcing offline accuracy constraints.

As shown in Figure~\ref{fig:timeline}, \name{} adopts a pipelined execution model where GPU-side attention and FFN for layer~$i$ (the \textit{GPU stage}) overlap with CPU-side logit computation and data transfer for layer~$i{+}1$ (the \textit{CPU stage}). The scheduler monitors the execution time of both stages and 
keeps the CPU stage hidden behind the GPU stage. At runtime, it controls two parameters: the number of 
tokens ($K$) processed on the CPU, and the corresponding token selection strategy. 

\subsubsection{Adaptive Selection of Token Count ($K$)}
\name{} initially applies \textit{Post-QK$^{\mathrm{T}}$ selection} (Figure~\ref{fig:token_selection}), where the CPU computes attention logits for all offloaded KV tokens and selects the top-$K$ most important ones. This is accuracy-friendly as importance is derived from the full context. The feedback scheduler increases $K$ as long as the combined CPU computation and transfer latency remains below the GPU-stage latency, thereby maximizing accuracy without stalling the pipeline; otherwise, it decreases $K$ to reduce overhead. 

\begin{figure}[t]
\centering
\subfloat[OPT-6.7B.\label{fig:acc_opt_6_7b}]{
    \centering\includegraphics[width=0.48\linewidth]{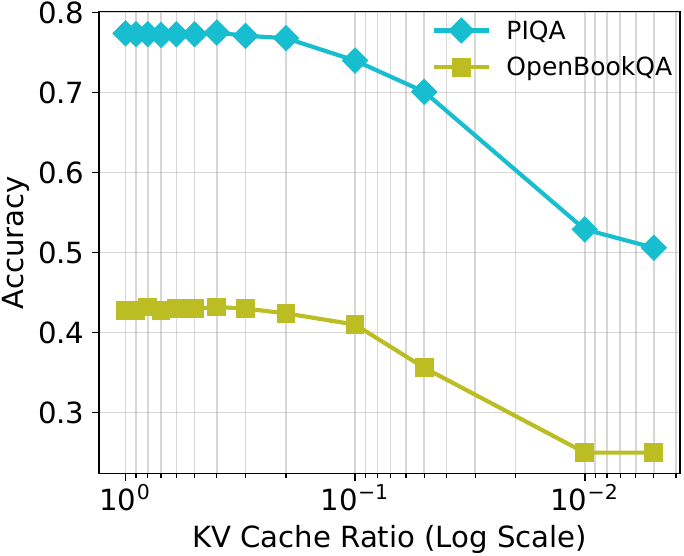}
}
\subfloat[OPT-13B.\label{fig:acc_opt_13b}]{
    \centering\includegraphics[width=0.48\linewidth]{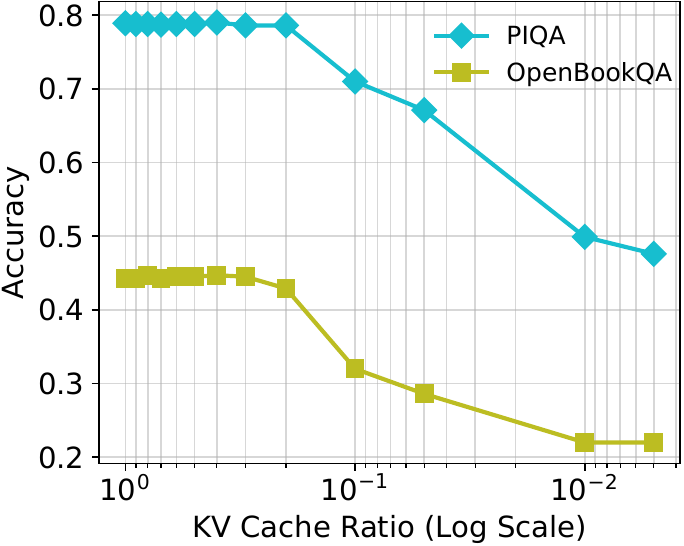}
}
\vspace{-5pt}
\caption{Accuracy of OPT-6.7B and OPT-13B on PIQA and OpenBookQA datasets when retaining different percentages of historical token. 
}
\label{fig:accuarcy_diff_kv}\vspace{-5pt}
\end{figure}

To preserve output quality, the scheduler enforces a lower bound $K_{\min}$ determined offline through accuracy profiling. As illustrated in Figure~\ref{fig:accuarcy_diff_kv}, representative models exhibit stable accuracy inflection points across datasets, enabling identification of a robust, model-specific $K_{\min}$. 
Unlike prior KV cache pruning methods~\cite{infinigen2024lee,zhang2023h2o,xiao2024streaming,ghadia2025dialogue} that rely on a fixed $K$, \name{} dynamically adjusts $K$ at runtime while respecting this lower bound, \textit{to maximize performance while preserving accuracy}. A single $K$ value is used across layers, while the selected tokens may vary per layer according to the selection algorithm. 

\subsubsection{Dynamic Switching Between Post- and Pre-QK$^{\mathrm{T}}$ Selection}
As sequence length increases, Post-QK$^{\mathrm{T}}$ selection can become CPU-bound even when $K$ reaches $K_{\min}$. When the scheduler detects that CPU computation becomes the bottleneck, it switches to \textit{Pre-QK$^{\mathrm{T}}$ selection} (Figure~\ref{fig:token_selection}), where logits are computed only for the selected $K$ tokens. This significantly reduces CPU overhead and allows \name{} to sustain pipelined execution under large KV cache sizes. In this mode, important tokens are identified using lightweight prediction mechanisms (e.g., position-based heuristics), described in Section~\ref{sec:tokensel}.

Algorithm~\ref{alg:scheduler} summarizes the feedback scheduler policy, which dynamically adjusts $K$ and switches between Post- and Pre-QK$^{\mathrm{T}}$ selection based on runtime latencies while enforcing accuracy constraints.

\subsection{Semantic-Aware KV Mapping on Tiered Memory}
\label{sec:semantic}
\name{} is designed to scale beyond GPU memory and host DRAM capacity by leveraging tiered memory such as CXL. However, naively placing KV caches in CXL memory can significantly degrade performance due to its higher access latency. To address this, \name{} introduces a \textit{semantic-aware KV cache mapping} strategy that exploits the distinct roles of key ($\mathbf{K}$) and value ($\mathbf{V}$) vectors in attention computation.

A key observation is that $\mathbf{K}$ and $\mathbf{V}$ exhibit different access characteristics in hybrid attention. 
As described in Section~\ref{sec:attn_parallelization}, CPU-side computation in \name{} is restricted to attention logit (query--key dot products), accessing only $\mathbf{K}$ vectors, while $\mathbf{V}$ vectors are not used on the CPU and are later transferred to the GPU for aggregation.

Exploiting this semantic distinction, \name{} places \textit{$\mathbf{K}$ vectors} in CPU DRAM to provide faster accesses 
during CPU logit computation, while storing \textit{$\mathbf{V}$ vectors} in CXL 
memory 
once the KV cache exceeds DRAM capacity. This ensures that CPU execution avoids CXL access latency on the critical path. $\mathbf{V}$ vectors 
are transferred to GPU via DMA, allowing CXL to act as a capacity extension without incurring NUMA penalties for CPU execution. When DRAM is exhausted, $\mathbf{K}$ vectors are evicted using an LRG policy, consistent with Section~\ref{sec:gpucomp}.

Unlike prior methods relying on runtime access profiling or hotness tracking, \name{}'s semantic-aware mapping is determined by attention semantics and requires no online monitoring, providing stable and predictable performance.
Note that the existing access hotness-based mapping is ineffective for KV caches, as no tokens are consistently accessed more frequently across queries, which explains the needs for diverse token selection strategies, as discussed in Section~\ref{sec:tokensel}.


\subsection{Important Token Selection for Scalable CPU Performance}
\label{sec:tokensel}


To balance CPU and GPU performance, the feedback scheduler limits CPU-side attention computation by selecting which offloaded KV tokens participate in attention.
\name\ supports this via a unified \textit{Token Selector} module (Figure~\ref{fig:design}), enabling integration with diverse token selection algorithms.

Existing token selection methods fall into three categories: (1) position-based heuristics that exploit position patterns of important tokens~\cite{xiao2024streaming, ghadia2025dialogue, liu2023scissorhands}, (2) score-based methods that rank tokens using attention scores~\cite{infinigen2024lee, zhang2023h2o, li2024snapkv}, and (3) predictors that identify important attention heads~\cite{liu2023deja, feng2024ada}. All these approaches produce a ranked list of candidate tokens or heads according to estimated importance.

The token selector determines which tokens to process 
based on this ranked list, along with the token count $K$ and the selection strategy provided by the feedback scheduler. Under Post-QK$^{\mathrm{T}}$ selection,  it applies score-based methods to select the top-$K$ tokens for the logit calculator.
Under Pre-QK$^{\mathrm{T}}$ selection, it uses lightweight predictors (e.g., position- or head-based) to identify the top-$K$ tokens. In Section~\ref{sec:eval}, we evaluate \name{} with multiple token selection algorithms, demonstrating its general applicability across diverse selection strategies.

\subsection{Workflow}
Figure~\ref{fig:workflow} illustrates the execution workflow of \name{} during the generation (or decode) phase. 
\bcircled{1} While the GPU executes the attention and FFN of transformer layer $i$, the CPU simultaneously prepares computation for layer $i+1$ by selecting important 
tokens based on feedback scheduler decisions. 
If pre-QK$^{\mathrm{T}}$ selection is chosen, the token selector picks the top-$K$ tokens, and the CPU speculatively computes their attention logits using the query vector $Q$ from the previous layer.
If post-QK$^{\mathrm{T}}$ selection is chosen, logits are computed for all offloaded tokens and then filtered based on scheduler and selector inputs.
\bcircled{2}--\bcircled{3} After logit computation, the attention logits of selected tokens, along with their corresponding value vectors, are prefetched into GPU memory.
Since value vectors are not accessed on the CPU, they are transferred directly from their backing storage (e.g., CXL-expanded memory) to the GPU.
\bcircled{4} In parallel, the GPU computes attention logits for tokens already cached in GPU memory, typically recent tokens.
\bcircled{5} The GPU then concatenates CPU- and GPU-computed logits and value vectors, restoring the original token order via the token mapper to form a unified attention logit vector equivalent to monolithic execution.
Then, the GPU applies softmax over the concatenated logits, performs value aggregation, and completes the remaining computations of the transformer block (e.g., FFN). 
The resulting hidden states are passed to the next layer, and the pipeline repeats.
\bcircled{6} Finally, the GPU sends the transformer block output to the CPU for the next layer computation.

\subsection{Implementation} 
We developed \name{} on top of FlexGen~\cite{sheng2023flexgen}, a widely used KV-cache offloading inference framework, leveraging its KV cache management and data movement utilities. In its Python frontend, we decouple attention-logit computation from the rest of the attention pipeline, implement the feedback scheduler, and extend the prefetching logic to support CPU–GPU pipelined execution.
To support semantic-aware mapping on tiered memory, we modified PyTorch’s CPU allocator~\cite{pytroch_cpu_alloc} in its C++ backend using the Linux \texttt{mbind} API~\cite{mbind_manpage}.

\begin{figure*}[t]
\centering
\includegraphics[width=0.95\linewidth]{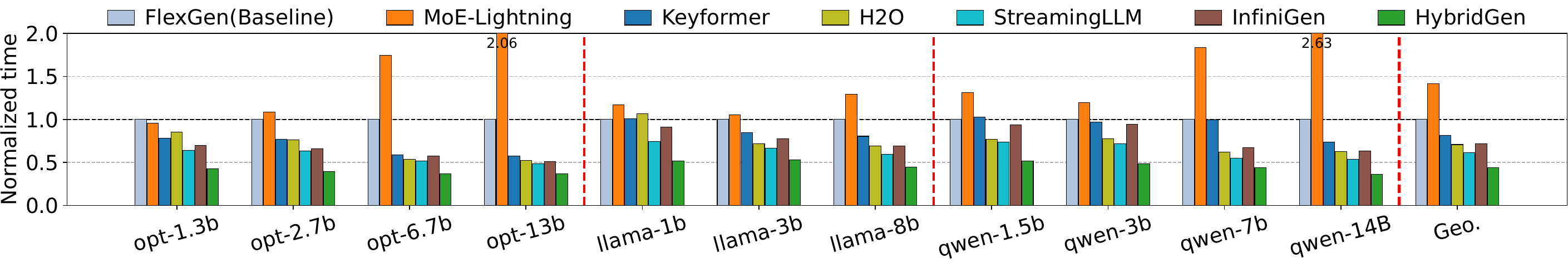}
\vspace{-5pt}
\caption{End-to-end latency of different models. (Normalized to Baseline)
}
\Description{}
\label{fig:model_time}\vspace{-10pt}
\end{figure*}

\subsection{Generalization to Other Frameworks}
\name{} is designed to be framework-agnostic and compatible with existing LLM serving systems. It relies on three capabilities commonly available in modern frameworks: (1) a decode-time attention path exposing the current hidden state or query vector, (2) a KV-cache manager that tracks logical token order across GPU and host memory, and (3) a runtime scheduler that overlaps CPU and GPU execution. These abstractions are already supported in systems such as vLLM and SGLang.
Integration primarily replaces monolithic decode-time attention with a split execution path. The CPU computes attention logits for offloaded tokens and gathers corresponding value vectors, while the GPU computes logits for GPU-resident tokens, merges results in logical token order, applies softmax, and completes value aggregation and remaining transformer operations. Other components (e.g., model loading, batching, sampling, prefix caching) remain unchanged. In paged KV-cache systems, integration requires extending KV metadata to track device residency and adding a lightweight feedback controller to adapt the number of CPU-processed tokens. If tiered memory is available, \name{} further applies semantic-aware placement by keeping key vectors in host memory and placing value vectors in lower tiers.

\section{Evaluation}\label{sec:eval}

\subsection{Experimental Setup}

\begin{table}[b]
\vspace{-10pt}
\caption{Hardware and Software Environment.}
\vspace{-5pt}
\label{tab:platform}\vspace{-5pt}
\centering
\scriptsize
\begin{adjustbox}{width=0.5\textwidth}
\begin{tabular}{|c|c|c|c|c|c|c|c|}
\hline
Machine & CPU & GPU & System &
\makecell{DRAM} &
\makecell{CXL\\Memory} &
\makecell{GPU\\Driver} &
\makecell{CUDA\\Toolkit} \\
\hline
\hline
A &
\makecell{Intel Xeon\\Gold 5320} &
\makecell{A100\\(80GB)} &
Linux 5.14 &
128 GB &
-- &
570.86.10 &
12.1 \\
\hline
B &
\makecell{Intel Xeon\\Gold 6530} &
\makecell{H100 NVL\\(94GB)} &
Linux 6.8 &
512 GB &
-- &
580.95.05 &
12.5 \\
\hline
\
C &
\makecell{Intel Xeon\\ 6960P} &
\makecell{GeForce\\RTX 5090\\(32GB)} &
\makecell{Linux 6.2.6} &
\makecell{128GB} &
\makecell{128 GB\\(DDR4-3200\\PCIe 5.0)} &
570.86.10 &
12.8 \\
\hline
\end{tabular}\vspace{-5pt}
\end{adjustbox}
\vspace{-5pt}
\end{table}

\subsubsection{Workloads and Platforms.}
We evaluate a diverse set of LLMs, including OPT\cite{zhang2022opt} (1.3B–13B), Llama-3.2~\cite{grattafiori2024llama} (1B and 3B), Llama-3.1~\cite{grattafiori2024llama} (7B), and Qwen2.5~\cite{qwen2.5} (1.5B–14B).
Experiments are conducted on three platforms with different CPU–GPU configurations and memory hierarchies, as summarized in Table~\ref{tab:platform}. Machine~A is used as the primary evaluation platform, Machine~B for sensitivity analysis across GPU generations, and Machine~C to evaluate \name{} on a CXL-enabled system. 
We use OPT models for most evaluations to enable comparison with prior KV-cache optimization work. For long-context experiments, we use Llama and Qwen models, since OPT supports sequences only up to 2K.
Unless otherwise specified, the default sequence length is 2K and the batch size is 1.

\subsubsection{Baseline \& SOTA Solutions.}
We compare \name{} against six SOTA KV cache management methods, categorized by attention computation affinity and token selection strategy. We distinguish full-attention methods (AoG and AoC)  
from selective attentions that perform token selection before (Pre-QK$^{\mathrm{T}}$) or after (Post-QK$^{\mathrm{T}}$) attention logit computation. Selective attention 
methods run attention on GPU.

\begin{enumerate}[label=\textbullet,leftmargin=*, labelindent=2pt]

\item \textbf{FlexGen} (AoG, Baseline)~\cite{sheng2023flexgen} offloads KV cache to lower-tier memory with zig-zag scheduling while keeping attention on the GPU. We offload 100\% of the KV cache and use it as the primary baseline.

\item \textbf{MoE-Lightning} (AoC)~\cite{moelightning}, originally designed for MoE models, offloads KV cache and attention to the CPU when GPU memory is exhausted, while the GPU executes FFN layers, reducing KV transfer but becoming CPU-bound at long sequences. 

\item \textbf{Keyformer} (Post-QK$^{\mathrm{T}}$)~\cite{adnan2024keyformer} selects top-$K$ tokens based on accumulated attention scores after attention computation and serves as our default Post-QK$^{\mathrm{T}}$ method.

\item \textbf{H2O} (Post-QK$^{\mathrm{T}}$)~\cite{zhang2023h2o} keeps heavy-hitter and recent tokens in a fixed-size KV cache, evicting less important entries, but still materializes attention logits over the retained cache on the GPU.

\item \textbf{StreamingLLM} (Pre-QK$^{\mathrm{T}}$)~\cite{xiao2024streaming}
uses attention sinks and a sliding window for static KV retention, enabling Pre-QK$^{\mathrm{T}}$ execution but potentially degrading long-range accuracy. We retain 4 sink tokens and a 1024-token window.

\item \textbf{InfiniGen} (Pre-QK$^{\mathrm{T}}$)~\cite{infinigen2024lee}
predicts important tokens using lightweight GPU-side approximations, but selection lies on the critical path, limiting efficiency when KV caches are largely offloaded. 
\end{enumerate}

\subsection{Performance}
\subsubsection{Overall Performance}

Figure~\ref{fig:model_time} shows end-to-end generation performance across all tested models. 
\name{} consistently achieves the best performance across all model scales. On average, it delivers up to 1.41$\times$ speedup over the state-of-the-art Pre-QK$^{\mathrm{T}}$ method (InfiniGen) and 1.86$\times$ over the Post-QK$^{\mathrm{T}}$ method (Keyformer), and up to 3.2$\times$ over full-attention and offloading baselines (FlexGen, MoE-Lightning).
These gains stem from \name{}'s hybrid execution design, which overlaps CPU-side attention logit computation with GPU execution while dynamically adjusting the number of selected tokens to avoid pipeline stalls. Compared to Post-QK$^{\mathrm{T}}$ approaches that run attention only on the GPU, \name{} reduces both attention computation and CPU--GPU data movement. Compared to Pre-QK$^{\mathrm{T}}$ methods with static selection, it adapts token selection at runtime, improving accuracy while maintaining balanced and high attention throughput. Accuracy results are in Section~\ref{sec:accuracy_eval}. 

\subsubsection{\textbf{Performance with Different Sequence Length.}}
\begin{figure}[t]
\centering
\subfloat[\textbf{Qwen2.5-7B} \label{fig:diff_len_time_qwen}]{
\centering\includegraphics[width=0.9\linewidth]{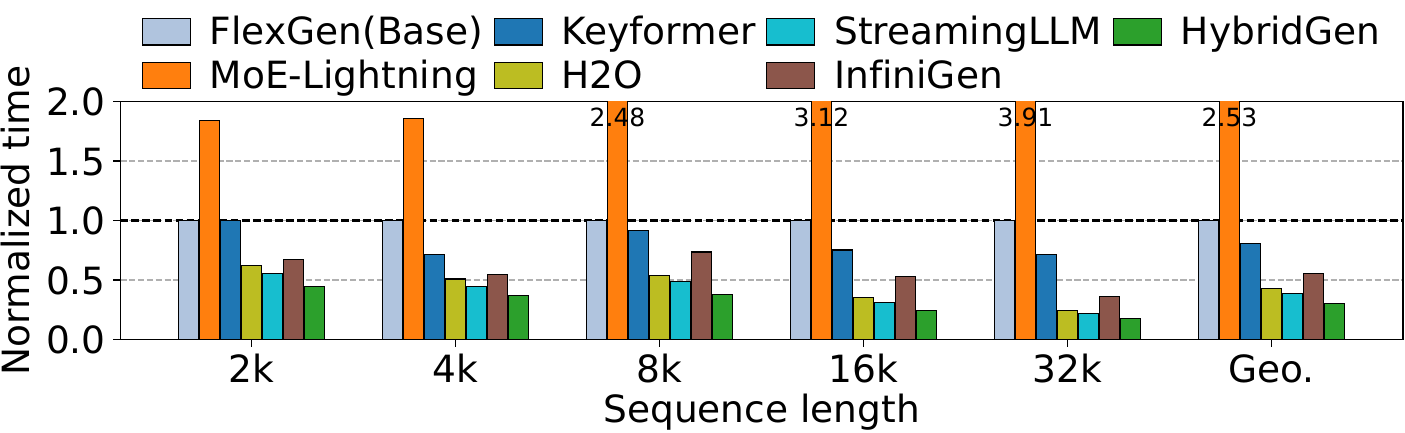}
}\vspace{-5pt}
\\
\subfloat[\textbf{Llama-3.1-8B} \label{fig:diff_len_time_llama}]{
\centering\includegraphics[width=0.9\linewidth]{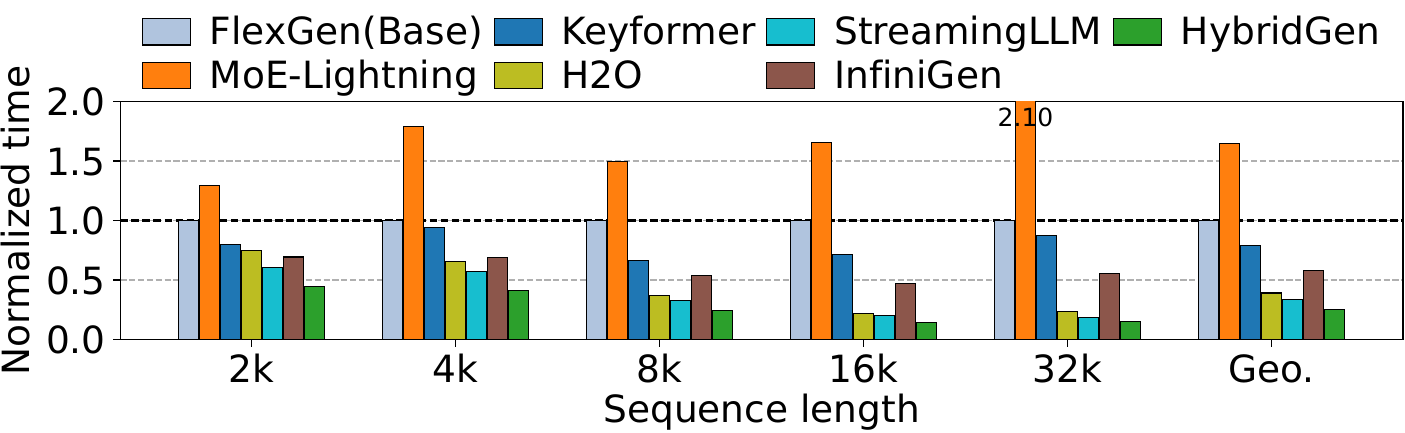}
}\vspace{-10pt}
\caption{Performance under different sequence length.}
\Description{}
\label{fig:diff_len_time_long}
\end{figure}

Figure~\ref{fig:diff_len_time_long} reports the end-to-end generation latency as sequence length increases from 2K to 32K for Qwen2.5-7B and Llama-3.1-8B. 
\name\ consistently achieves the best performance across all sequence lengths and shows even better speedup than baseline for longer sequences. 
On average, \name\ reduces latency by up to 1.82$\times$ and 2.34$\times$ over pre-QK$^{\mathrm{T}}$ methods, and by up to 2.66$\times$ and 3.18$\times$ over post-QK$^{\mathrm{T}}$ methods for Qwen2.5-7B and Llama-3.1-8B, respectively. %
This benefit stems from \name’s runtime load balancing, which bounds GPU attention cost by offloading attention logit computation to the CPU and dynamically adjusts the amount of CPU computations 
to avoid pipeline stalls. Note that the accuracy is preserved by processing at least $K_{\min}$ tokens.
In contrast, pre-QK$^{\mathrm{T}}$ baselines incur increasing GPU-side prediction or attention overhead as sequence length grows, while post-QK$^{\mathrm{T}}$ methods continue to perform attention over a large token set.

\subsubsection{\textbf{Performance with Different Batch Size.}}
\begin{figure}[t]
\centering
\includegraphics[width=0.9\linewidth]{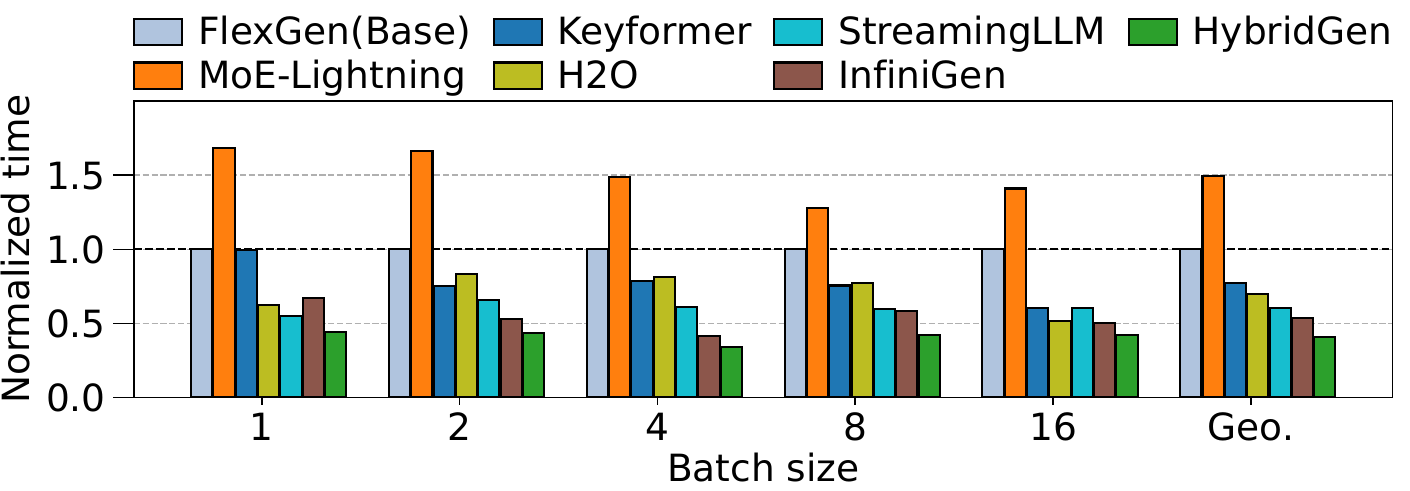}\vspace{-10pt}
\caption{Performance 
under different batch size.}
\Description{}
\label{fig:batch_size_time}\vspace{-10pt}
\end{figure}

Figure~\ref{fig:batch_size_time} shows end-to-end generation latency across different batch sizes for OPT-13B. 
\name\ consistently achieves the best speedup 
across all batch sizes, and its performance advantage grows with larger batches.
On average, \name\ provides up to 1.5$\times$ speedup over the pre-QK$^{\mathrm{T}}$ methods 
and up to 1.87$\times$ speedup over post-QK$^{\mathrm{T}}$ approaches. 
This improvement is attributed to \name’s ability to overlap CPU-side attention logit computation with GPU execution and dynamically adjust the number of selected tokens to balance CPU--GPU performance. 
In contrast, post-QK$^{\mathrm{T}}$ methods suffer from increased GPU attention cost as batch size grows, while pre-QK$^{\mathrm{T}}$ baselines with static token retention exhibit limited scalability due to fixed attention windows or GPU-side prediction overhead. 

\subsubsection{\textbf{Impact of Semantic-Aware KV Cache Mapping.}}
\begin{figure}[t]
\centering
\includegraphics[width=0.9\linewidth]{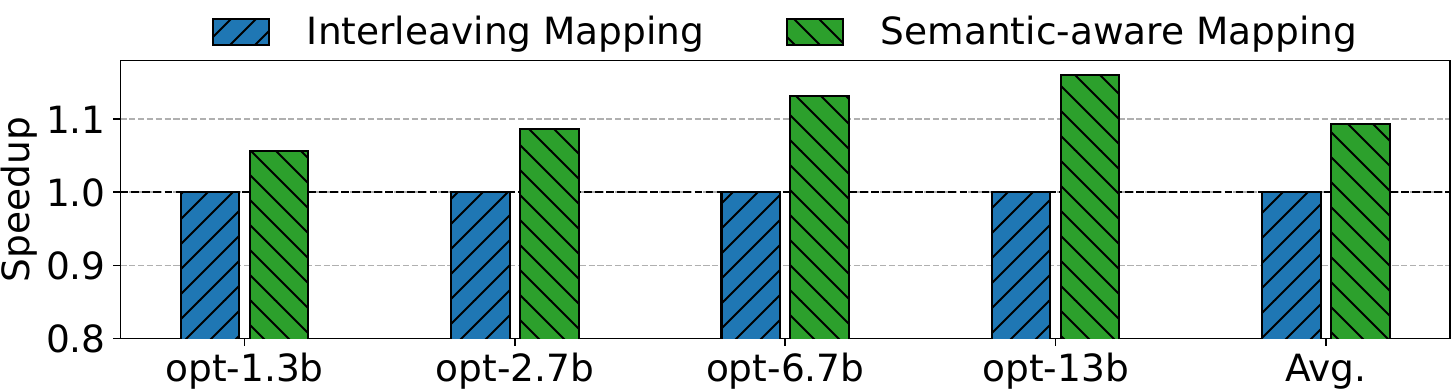}\vspace{-10pt}
\caption{Speedup of semantic-aware mapping.}
\Description{}
\label{fig:semantic_mapping}\vspace{-5pt}
\end{figure}

Figure~\ref{fig:semantic_mapping} evaluates the impact of \name{}’s semantic-aware KV mapping. 
We compare it with a widely used page mapping, which interleaves 
pages across local DRAM and CXL memory~\cite{ipdps25:cxl, sun2023cxl, zhao2025optimizing}. Semantic-aware mapping consistently outperforms
, achieving increasing speedups for larger models. 
This trend reflects the growing significance of data 
placement for larger KV caches. 
The speedup 
stems from \name{}’s faster $\mathbf{K}$ vector accesses for the attention logit computations, while $\mathbf{V}$ vectors can be transferred directly from the CXL memory to the GPU via DMA. 
In contrast, the interleaving mapping incurs frequent accesses to CXL memory during CPU-side logit computation, causing longer latency on the critical path. 
These results show that semantic-aware KV mapping is essential for scalable hybrid attention on tiered memory.

\begin{figure}[t]
\centering
\includegraphics[width=0.9\linewidth]{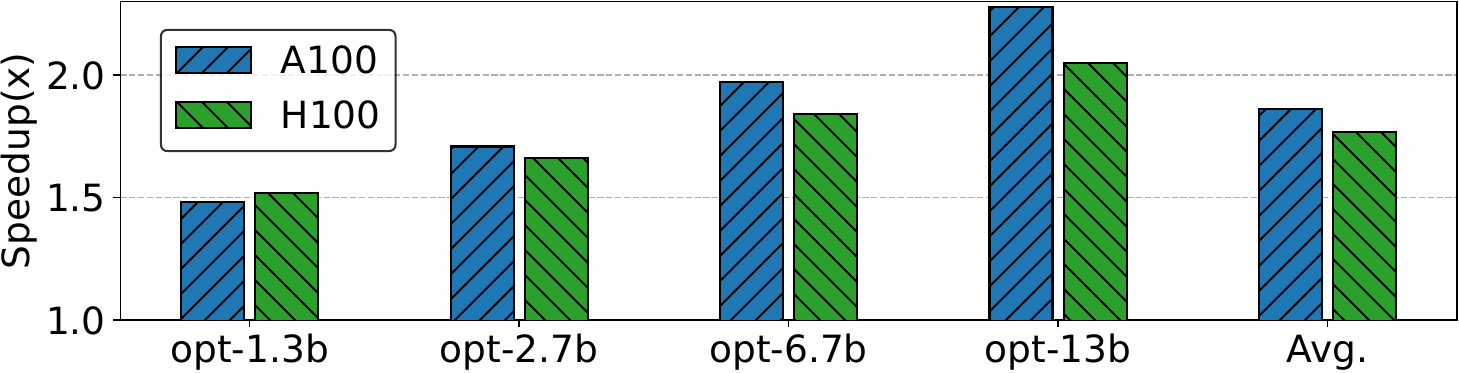}\vspace{-10pt}
\caption{Speedup of \name\ on different GPUs.} 
\Description{}
\label{fig:diff_gpu_perf}\vspace{-5pt}
\end{figure}

\subsubsection{Impact of GPU Architectures.}
\label{sec:sensitivity_gpu}

We evaluate \name{} on two GPU platforms (machines A and B in Table~\ref{tab:platform}) 
to understand \name's general effectiveness 
across GPU generations.
As shown in Figure~\ref{fig:diff_gpu_perf}, across all model sizes, \name{} consistently outperforms the 
baseline on both platforms, with higher 
throughput on H100 due to increased GPU compute capability and faster CPU--GPU interconnect.
The speedup trends remain stable across architectures, which proves the general effectiveness of \name. 

\begin{figure}[t]
\centering
\includegraphics[width=0.9\linewidth]{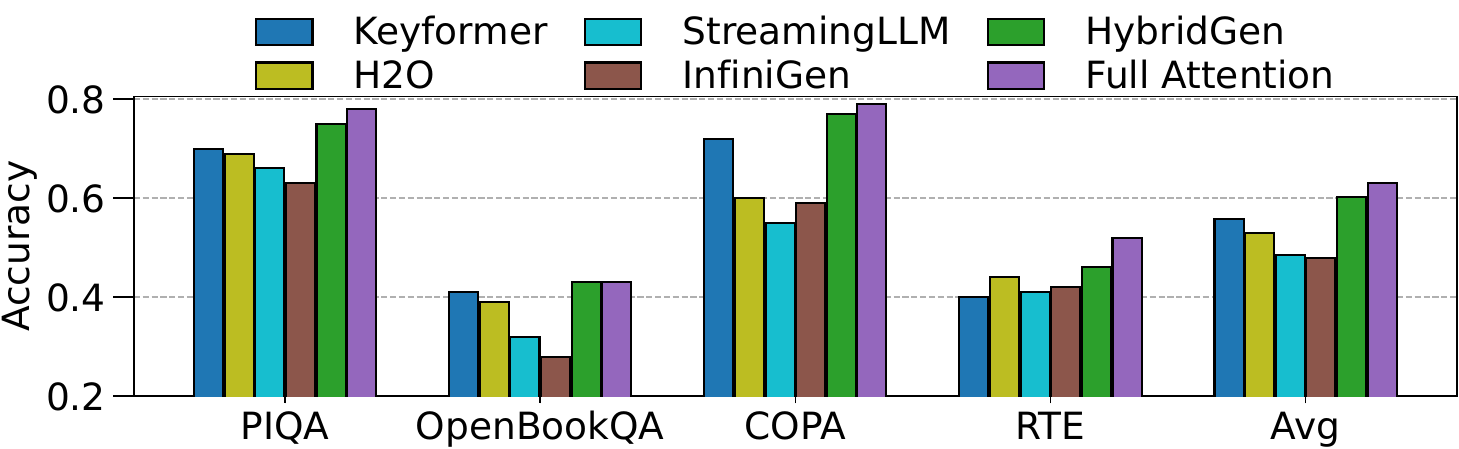}
\vspace{-10pt}
\caption{Accuracy 
under different KV cache pruning.}
\Description{}
\label{fig:opt_accuracy}\vspace{-10pt}
\end{figure}
\subsection{Accuracy}\label{sec:accuracy_eval}


We evaluate accuracy 
using the LM Evaluation Harness~\cite{gao2021framework} with a 3-shot prompting setup. 
Following prior work~\cite{biderman2024lessons, dutta2024accuracy, gao2021framework}, we report task-specific accuracy as the primary metric, measured by exact match or multiple-choice accuracy depending on the dataset. All methods use identical prompts and decoding configurations, and results are averaged over all evaluation samples to ensure fair comparison.
We evaluate on four benchmarks: PIQA~\cite{bisk2020piqa}, OpenBookQA~\cite{mihaylov2018can}, COPA~\cite{roemmele2011choice}, and RTE~\cite{wang2018glue}.
Figure~\ref{fig:opt_accuracy} shows the results. 
Compared to the full attention, which derives the best accuracy by preserving the complete context, 
\name\ closely matches the accuracy 
across all datasets, with only an average accuracy gap of 0.02 across datasets. 
This small gap proves the effectiveness of \name's feedback-driven token selection with $K_{\min}$ threshold.
In contrast, baselines with static or aggressive token selection (StreamingLLM and InfiniGen) exhibit significant accuracy loss, particularly on tasks that require long-range reasoning (e.g., COPA and RTE).
Post-QK$^{\mathrm{T}}$ approaches 
retain more accuracy than static pre-QK$^{\mathrm{T}}$ methods but still degrade due to limited retained context.

\subsection{\textbf{Effectiveness of the Feedback Scheduler}}

\begin{figure}[t]
\centering
\includegraphics[width=0.9\linewidth]{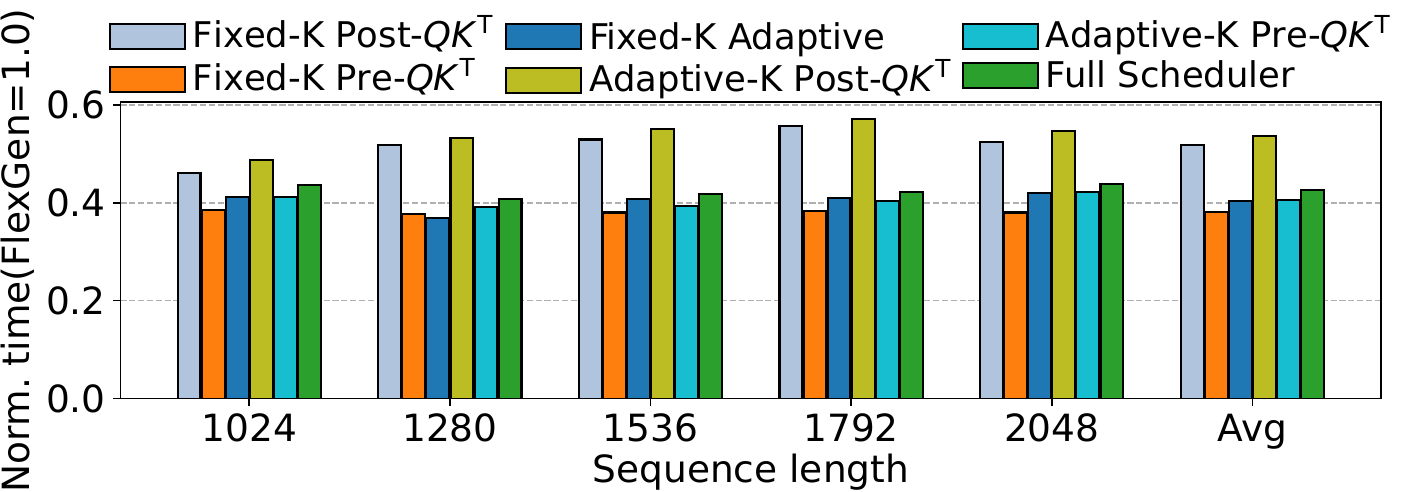}
\vspace{-5pt}
\caption{Performance 
under different feedback scheduler configurations across varying sequence lengths.}
\Description{}
\label{fig:feedback_perf}\vspace{-10pt}
\end{figure}

\begin{figure}[t]
\centering
\includegraphics[width=0.9\linewidth]{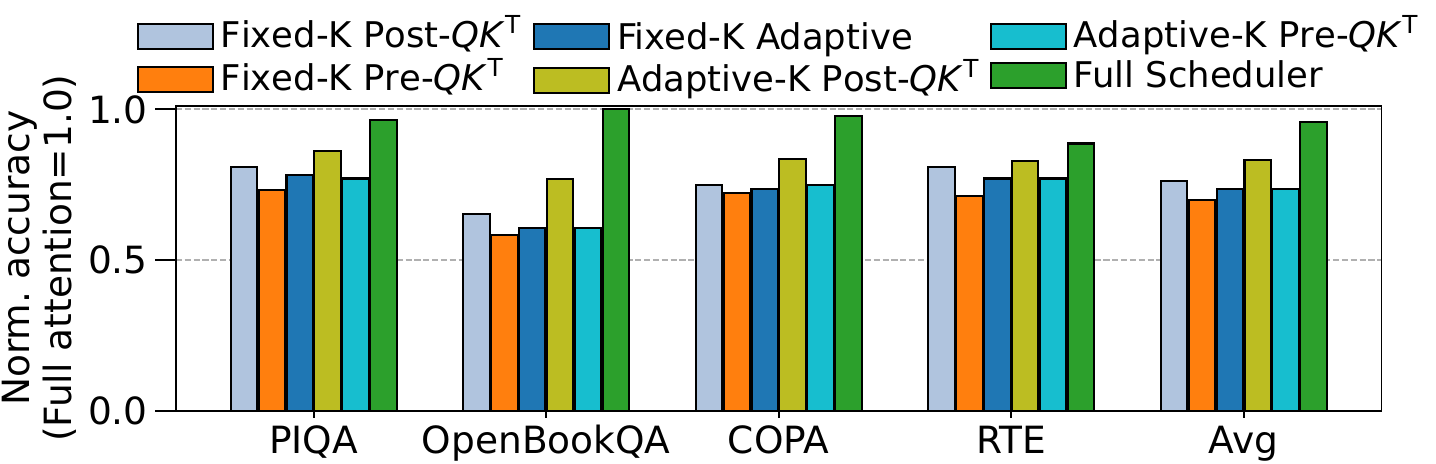}
\caption{
Accuracy 
under different feedback scheduler configurations across varying datasets.}
\Description{}
\label{fig:feedback_accu}\vspace{-10pt}
\end{figure}

To understand the impact of feedback scheduler, we measured performance and accuracy impact under different feedback scheduler configurations by varying the token selection timing (Pre- vs.\ Post-QK$^{\mathrm{T}}$) and token count control ($K$).
Figure~\ref{fig:feedback_perf} and Figure~\ref{fig:feedback_accu} show the results.
When a fixed $K$ is used (e.g., 128, which is $K_{min}$ of OPT-13B), as shown in the first three bars, Pre-QK$^{\mathrm{T}}$ and Post-QK$^{\mathrm{T}}$ respectively show the best and the worst performance, as Pre-QK$^{\mathrm{T}}$ does not compute attention scores for token selection, while it is required by Post-QK$^{\mathrm{T}}$. When the selection timing is dynamically controlled based on performance (\textit{Fixed-K Adaptive}), the performance is almost close to the Pre-QK$^{\mathrm{T}}$ with marginal slowdown due to the added computations when Post-QK$^{\mathrm{T}}$ is chosen. However, as Post-QK$^{\mathrm{T}}$ can select important tokens more accurately, the adaptive approach derives better accuracy than Pre-QK$^{\mathrm{T}}$, as shown in Figure~\ref{fig:feedback_accu}. 

When $K$ is selected adaptively (the last three bars), the speedup is almost similar to fixed $K$ configurations. A slight slowdown is observed due to the longer execution time when a larger $K$ is used. Note that we used the minimum $K$ ($K_{min}$) for fixed $K$ configurations; the scheduler increases $K$ when CPU has room to process more tokens. However, as a longer context can be considered when a bigger $K$ is chosen, the adaptive $K$ selection shows superior accuracy than fixed $K$ configurations. When $K$ and selection timing are both chosen at runtime (\textit{Full Scheduler}), the accuracy becomes significantly higher than the other methods while the performance is maintained similarly to the Pre-QK$^{\mathrm{T}}$. 

Together, these results demonstrate that the feedback scheduler enables \name{} to achieve an effective balance between performance and accuracy.

\section{Related Work}

\textbf{KV Cache Memory Management:}
A range of systemic and algorithmic techniques have been proposed to manage the growing KV cache footprint in LLM inference. vLLM~\cite{kwon2023efficient} partitions the KV cache into fixed-size blocks, enabling non-contiguous storage. vAttention~\cite{prabhu2025vattention} decouples virtual and physical KV cache allocation using CUDA virtual memory APIs.
Tetris~\cite{zhang2025jenga} addresses non-uniform memory pressure across layers using an attention-aware allocator to improve GPU memory utilization.
LMPrefill~\cite{du2025prefillonly} targets prefill-only inference by storing only the final layer’s KV cache.

\textbf{Multi-Tier KV Cache:} To alleviate the memory pressure,
prior work explores KV cache offloading across memory tiers. 
FlexGen~\cite{sheng2023flexgen} offloads tensors (e.g., KV cache) across GPU, CPU, and disk with optimized placement and scheduling.
FlashGen~\cite{jeong2025accel} proposes a multi-level KV cache spanning GPU, CPU, and SSD with request reordering to improve utilization. Pensieve~\cite{yu2025stateful} introduces a multi-tier KV cache for multi-turn conversations across GPU and CPU memory.

\textbf{KV Cache Pruning/Compression:} 
To reduce data movement, KV cache pruning and compression techniques have been proposed.
InfiniGen~\cite{infinigen2024lee} uses rehearsal-based token selection and prefetching to reduce KV transfer overhead. 
MorphKV~\cite{ghadia2025dialogue} maintains a fixed-size KV cache by selecting important tokens based on attention scores.

\textbf{Unlike these solutions, \name\ provides a unified framework for hybrid attention execution in long-context LLM inference.} 
It enables CPU and GPU to collaboratively compute attention over locally resident KV tokens, dynamically balances workloads, and introduces semantic-aware KV mapping for efficient use of CXL-expanded memory.
\section{Conclusion}
This paper presents \name\, a new hybrid attention framework for long-context LLM inference. \name\ aims at maximizing the utilization of underlying hardware resources, which not only improves performance but also reduces data transfers by allowing both CPU and GPU to compute attention with locally stored data. Our proposed intra- and inter-layer attention dependency resolutions, feedback scheduler, and semantic-aware data mapping enforce scalable LLM inference by dynamically balancing the performance of CPU and GPU and leveraging larger CXL-expanded memory without NUMA overhead. The experiments show that \name\ outperforms state-of-the-art solutions by 1.41$\times$--3.2$\times$ on average.

\bibliographystyle{configs/ACM-Reference-Format}
\bibliography{references}

\end{document}